# Multiple spatial frequencies wavefront sensing

Roberto Ragazzoni[a,c], Daniele Vassallo[a,b,c], Marco Dima[a,c], Elisa Portaluri[a,c], Maria Bergomi[a,c], Davide Greggio[a,c], Valentina Viotto[a,c] Federico Biondi[a,c], Elena Carolo[a,c], Simonetta Chinellato[a,c], Jacopo Farinato[a,c], Demetrio Magrin[a,c], Luca Marafatto[a,c]

[a] INAF - Osservatorio Astronomico di Padova, Vicolo dell'Osservatorio 5, 35122 Padova, Italy
[b] Dipartimento di Fisica ed Astronomia - Università degli Studi di Padova, Vicolo dell'Osservatorio 3, 35122 Padova, Italy
[c] ADONI – Laboratorio nazionale per l'Ottica Adattiva, Italy

## ABSTRACT

We describe the concept of splitting spatial frequency perturbations into some kind of pupil planes wavefront sensors. Further to the existing approach of dropping higher spatial frequency to suppress aliasing effects (the so-called spatial filtered Shack-Hartmann), we point out that spatial frequencies splitting and mixing of these in a proper manner, could be handled in order to exhibit some practical or fundamental advantages. In this framework we describe the idea behind such class of concepts and we derive the relationship useful to determine if, by which extent, and under what kind of merit function, these devices can overperform existing conventional sensors.

## 1. INTRODUCTION

The pyramid[1] WaveFront Sensor (WFS hereafter) is a kind of pupil-plane WFS characterized by a larger sensitivity[2,3] than other conventional WFS in use in Adaptive Optics (AO hereafter). While this is now widely recognized[4,5], there are constraints such that conventional WFSs can still exhibit comparable results[6] and one should properly consider if this class of WFS has reached a sort of fundamental limit or at least how much further degree of freedom is available to further improve such a design. We recently introduced the concept of Dark-WFS[7,8,9,10] that, although as a concept does not really indicate a certain specific optical layout but more generically a certain behavior, it should be classified as an even more generic class of solutions where photons are selected and conveyed in a manner that hopefully maximize the overall efficiency. In this respect one should see with a different perspective the hierarchical WFS[11] where one splits the whole collected light to different kinds of samplings. A revisitation of the behavior of the Pyramid WFS, especially taking into account its inherent Fourier filtering in the focal plane introduced by the four adjacent faces of the pyramid[12,13] especially in its non-modulated version [14], a solution well proven in the sky[15] leading to reconsider the non-modulated pyramid as a sort of Dark-WFS. While hereafter we trace the basis and we offer some suggestions on this theme, the interested reader should take a look at a more detailed paper[16] on this topic.

## 2. GEOMETRICAL VS. FOURIER DESCRIPTION

In geometrical approximation, when a perfectly flat wavefront is approaching the pin of the pyramid used in a non-modulated pyramid WFS (or simply a pyramid WFS), the light is split into four exact parts and illuminates uniformly and completely the four reimaged pupils. In Fourier approximation, however, the light is basically diffracted from the edges and will illuminate just the edges of the pupils in a manner that the superposition of the four pupil images will exhibit a roughly annular edge around the pupil, while for each of the four pupils the edges will exhibit non uniform illuminations depending upon which is their position with respect to the axes defining the edges of the pyramid. Of course any imperfection, both in the wavefront itself and in the pyramid manufacturing will make more light falling in the body of the pupils rather than on the edges. Central obstructions have here the role to further complicate the description, but are not considered at such a pictorial description level. Turned edges will make more and more light going into the inner portion of the pupils as well, and a roof-like imperfect pin of the pyramid will make a couple of pupils brighter than the other. However, under ideal conditions, and assuming one is sampling carefully the inner portion of the pupil neglecting the edges, pyramid-WFS is exhibiting a sort of dark-WFS behavior.

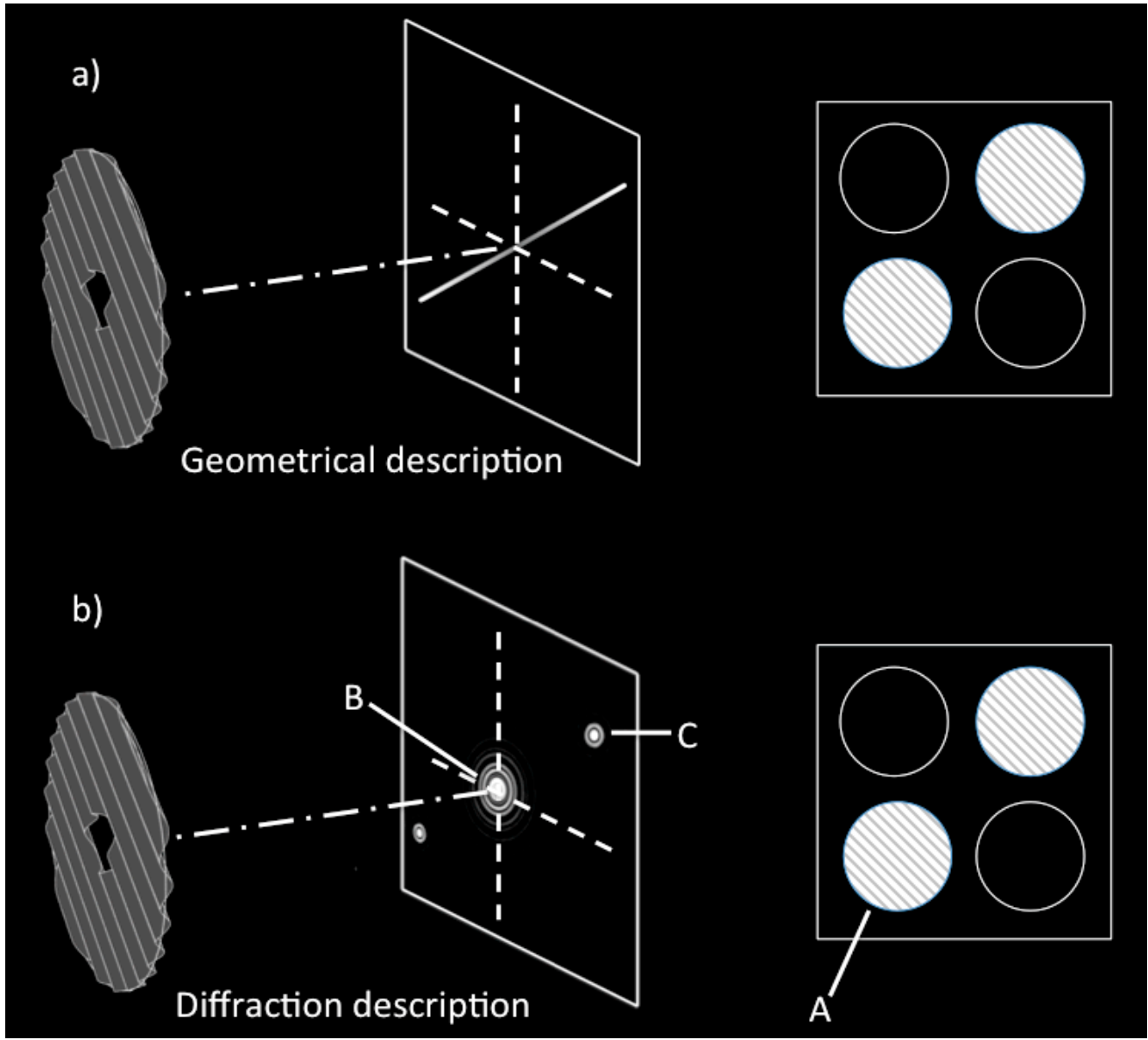

**Figure 1 A cylindrical shaped wavefront is incident on a pyramid WFS and its effects in the pupil plane (where the pyramid is located) and in the pupil plane (where the detector is located) are shown in a) geometrical approximation and b) under Fourier analysis properly taking into account the supposedly monochromatic incident light. It is to be noted that the fringe pattern occurring for instance in the pupil A is basically giving by the interference between the two point sources B and C that are able to illuminate such a region of the WFS detector.**

To further investigate the difference in the description of the pyramid under geometrical approximation vs. a Fourier analysis (that is to be considered here as representative of a non-approximation approach) we introduce in the wavefront a single frequency in the Fourier space domain. Basically this is represented by a wavefront that exhibits over its domain (for a real astronomical telescope an obstructed circular one) a cylindrical sine wave. In geometrical approximation a couple of pupils are basically left dark while the other two are showing a fringe pattern behavior given by the portion of the light that is falling away from the line of sight. Because of the sinewave the illumination on the focal plane is more intense at the edge of the extent where the light is distributed, as in the parts of higher slopes, both negative and positive, such a figure is slowly changing and occupies a significant fraction of the wavefront surface, in contrast with the peaks (both negative and positive) of the sine wave itself. In the fourier approximation, on the other hand, the focal plane is going to be illuminated just by two additional speckles, Airy-like shaped, that will occur at a distance from the line of sight that is only depend upon the spatial frequency of the sine wave perturbation, while their intensity will depends upon the overall rms, and by consequences by the amplitude, of the perturbed wavefront. In contrast the extension of the light on the focal plane in geometrical approximation will depend upon the value of the maximum first derivative, a quantity that depends in a multiplicative manner by both the spatial frequency and the amplitude of the perturbation. The fringe pattern on one of the pupils, finally, is given basically by the interference of the central and one of the now appeared lateral spot that acting as coherent point sources will produce a fringing effect with the proper spatial frequency in the pupil plane. Fig.1 collects all these considerations in a cartoon-style drawing and the reader should pay attention to it for a good comprehension of the phenomenon described so far. One of the interesting results is that, although the way the light is distributed in the focal plane, and the way the light is formed onto the pupil planes is radically different from qualitative and quantitative aspects, the final prediction is strikingly extremely coherent between the two approaches.

## 3. "DARKENING" A PYRAMID WFS

Following the Dark-WFS recipes all the photons that do not contribute to the final AO performances are basically better to be thrown away, otherwise their presence can only introduce additional Poissonian or photon shot noise. Following the same principles used in the Shack-Hartmann[13] in the case of the pyramid this can be clearly seen by an inspection of Fig.2. In principle such a selection can sharply remove all the light outside a certain region whose only influence is to illuminate the inner side of the pupils and to contribute somehow to the photon shot noise and hence to the measurements (and the limiting magnitude) of the WFS itself.

It is to be recalled that going to fainter magnitudes one would like to reduce the number of compensated modes such that the diaphragm properly used will probably need to be sized accordingly to the magnitude of the star under operation. Furthermore the selection of the modes not necessarily would cover the Fourier bidimensional space in a radially homogenous mode. We do not speculate here, although it deserves a mention, that under some specific circumstances (one could envisage, for instance, a DM with a squared grid of actuators) the diaphragm could lack of rotational symmetry and exhibits a square or even more specific and exotic shape.

Another consideration is that a sharp edge diaphragm in the focal plane would not –strictly speaking- cancel out spatial frequencies above a certain figure in a sharp way, as the speckles that appear on the focal plane do always have a finite size. It is to be investigated properly if an oversizing of the diaphragm by such a quantity, or a mask incorporating sort of apodizing profile progressively fades out the proper range of high frequencies.

On the practical side one is faced with the issue of nevertheless providing some sort of modulation for calibration purpose, or because –although proper "darkening" sounds like it only works on a non-modulated pyramid one- best results are obtained with a modest modulation. In all these cases modulating on a pupil plane through a steering mirror or mounting the diaphragm directly onto the pyramid would make its edges fluctuating in the Fourier space (an option that could be investigated to ensure if it could be used as a form of apodization described so far) and a way to overcome this would be to have an oscillating pyramid whose pin lies in the hole left by a fixed mounted diaphragm. We do not further speculate on possible optomechanical arrangement for such solutions here.

The amount of gain in limiting magnitude obtained here can be worked out in various manners with different assumptions. In general, as the assumption is to make one of the noise sources to dominate over the other, this leads to a factors two of gain amounting to 0.75mag. This is just to be considered as a very rough estimation of the order of magnitude of the quantities involved and is nevertheless an interesting gain worthwhile to further investigate and to optimize.

Application of the same concept to other kinds of WFS (like the Smartt one) are likely to introduce a similar gain and should be treated as a general approach.

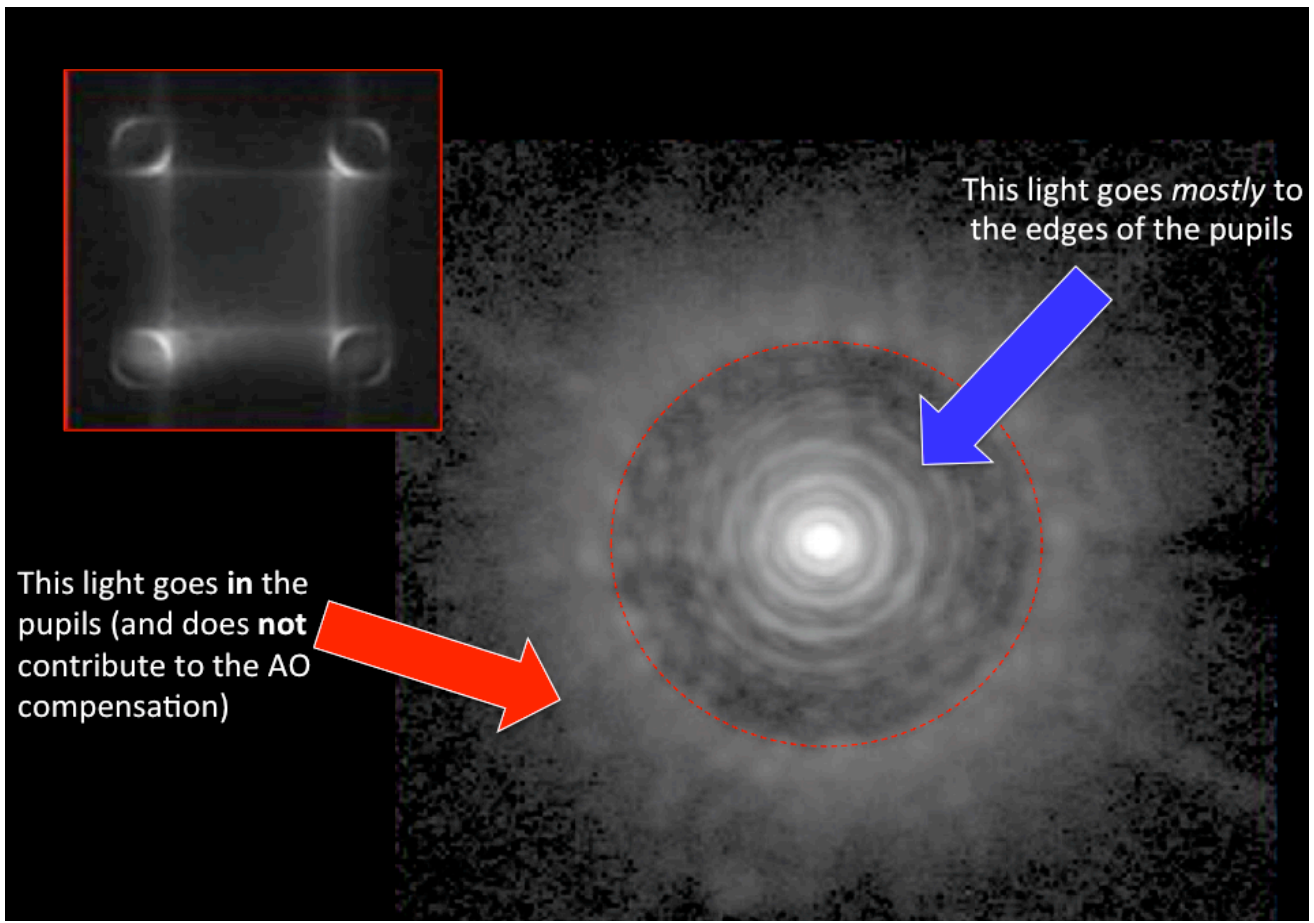

**Figure 2 The light outisde the so-called control region can only contribute to the noise in the pupil planes and are better to be thrown away. This can be done at almost zero cost leading to fainter limiting magnitude of the pyramid WFS.**

## 4. SELECTING SPATIAL FREQUENCIES

The concept described so far can be extended beyond the low-pass spatial filtering. In principle one could split the light into two (or more) spatial frequency regimes and use the selected light to sense the proper range of spatial frequencies with the caveat of not having the light coming from the uncompensated portion of the other spatial frequency range that introduces additional noise. As it is written this sounds like a way to sense aberrations in the most possible sensitive way. In fact, to some extent, focal plane wavefront sensing techniques, like the phase diversity, are characterized by pixels located specifically at a certain very limited spatial frequency range. Their inherent non-linearity and to some extent ambiguity in the phase reconstruction, made them far from this regime, however. Furthermore, recalling the way fringes in a pyramid WFS are generated, one should recall that isolating a certain range of spatial frequency without including the zero-th order one (the central peak in the PSF) will basically produce extremely weak signal, mainly from second order spikes in the PSF. This means that splitting does not favor a large number of regions in the spatial frequency domain to be selected, and even with just two the central peak light is to be split among the two regions in order to produce and efficient signal. At least in the pyramid and other similar pupil plane WFSs.

Selecting spatial frequency can be achieved not only to enhance the sensitivity of WFS, but to deliberately compensate only a certain range of them having, in exchange, some sort of gain. This approach could be the elusive answer to a kind of WFS that would be "loved" from spectroscopists leading to a PSF significantly larger than the diffraction limit but encompassing in a subarcsec region the vast majority of the collected light. Especially in large-z extragalactic astronomy this kind of AO would give large benefits to a potential high multiplexing instrument to look to radial velocity of remote galaxies.

In the past we looked at the same issue with the concept of a truncated pyramid, where the light is split into five beams each generating a pupil, where one is produced by the central region of a size comparable to the desired PSF size. Such a concept actually works in geometrical approximation as a ray will hit the outer four pupils only if the light gets out from the confined internal PSF region. Testing numerically however, it badly failed giving a very poor signal. The workaround for this is to include in this beam also the central part of the PSF with a spatially filtered beam producing the proper fringe patterns for every sampled spatial frequency (see also Fig.3).

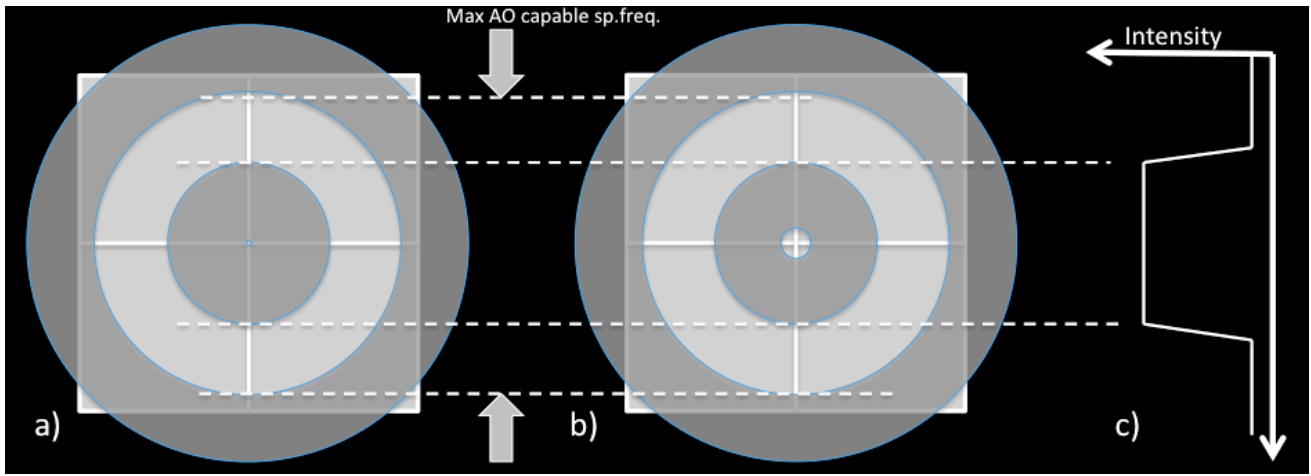

**Figure 3 In this pictorial description a pyramid WFS is being masked. For descriptive reason the full masking is shown with a small degree of tranparency to realize what is behind. From left to right: a) a concept working in geometrical approximation but failing to a detailed numerical simulation; b) the same but expected to work using the Fourier reasoning described in the text; c) the PSF that one is expected to generate for an AO system in closed loop with this kind of spatially frequency masked pyramid WFS.**

## 5. CONCLUSIONS

We described the extension of the spatial filtering concept to the pyramid WFS as a way to produce a device that works as close as possible to the dark-WFS concept. We extended this to selecting a certain range of spatial frequency and, in particular, we pointed out the possibility of masking in a way to engineer a PSF with a top-hat profile, that can be of some practical usefulness in astronomical extragalactic spectroscopy. As this is accomplished without the low order uncompensated mode light falling into the detector, at least to a large extent, this should come with a certain prize in terms of sensitivity, maybe leading to a significant increase in limiting magnitude for this new class of WFSs. The interested reader should also give a look to a specific paper[16] devoted to the same subject.